\documentclass[conference]{IEEEtran}
\IEEEoverridecommandlockouts
\usepackage[utf8]{inputenc}
\usepackage{cite}
\usepackage{floatrow}
\usepackage{amsmath,amssymb,amsfonts,stfloats}
\usepackage{tabularx}
\usepackage[left=0.63in,right=.63in,top=0.75in,bottom=1.1in]{geometry}
\usepackage{algorithmic}
\usepackage{graphicx}
\usepackage{textcomp}
\usepackage{floatrow}
\usepackage{xcolor}
\usepackage{booktabs}

\usepackage{multicol}
\usepackage{acronym}
\usepackage{url}
\usepackage{mathtools}
\usepackage[font=small]{caption}
\usepackage[caption=false,font=footnotesize]{subfig}
\usepackage{comment}
\def\BibTeX{{\rm B\kern-.05em{\sc i\kern-.025em b}\kern-.08em
    T\kern-.1667em\lower.7ex\hbox{E}\kern-.125emX}}

\newcommand{\linebreakand}{%
  \end{@IEEEauthorhalign}
  \hfill\mbox{}\par
  \mbox{}\hfill\begin{@IEEEauthorhalign}
}

\begin{document}

\title{ Multi-View Integrated Imaging and Communication
\thanks{This work was partially supported by the European Union under the Italian National Recovery and Resilience Plan (NRRP) of NextGenerationEU, partnership on “Telecommunications of the Future” (PE00000001 - program “RESTART”).}
}
\author{Davide~Tornielli~Bellini, Dario~Tagliaferri, Marouan~Mizmizi, Stefano~Tebaldini and Umberto~Spagnolini\\
Politecnico di Milano, Milan, Italy\\
E-mail (corresponding):\,name.surname@polimi.it}

\maketitle

\begin{abstract}
     Non-line-of-sight (NLOS) operation is one of the open issues to be solved for integrated sensing and communication (ISAC) systems to become a pillar of the future wireless infrastructure above 10 GHz. Existing NLOS countermeasures use either metallic mirrors, that are limited in coverage, or reconfigurable metasurfaces, that are limited in size due to cost. This paper focuses on integrated imaging and communication (IIAC) systems for NLOS exploration, where a base station (BS) serves the users while gathering a high-resolution image of the area. We exploit a large reflection plane, that is phase-configured in space and time jointly with a proper BS beam sweeping to provide a \textit{multi-view} observation of the area and maximizing the image resolution. Remarkably, we achieve a near-field imaging through successive far-field acquisitions, limiting the design complexity and cost. Numerical results prove the benefits of our proposal.
\end{abstract}


\section{Introduction}

Radio sensing is one of the paramount features of the future generation of wireless systems (6G) in the $3-300$ GHz band, to support novel market verticals~\cite{Wymeersch6G_ISAC}. Within the broad meaning of sensing, \textit{localization} is the procedure to estimate the position, velocity and possibly orientation of targets~\cite{10287134}, while \textit{imaging} refers to the generation of a map of the reflectivity of the environment, from which to infer the number of targets (via detection) and their shape \cite{tagliaferri2023cooperative}. Imaging is the first step in environment sensing, when no a-priori information on the environment is available. Localization follows after detection. The main key performance indicator used in localization is the estimation accuracy, with suitable lower-bounds \cite{Chetty2022_CRB}, while imaging quality is mostly described by \textit{resolution}, namely the capability of distinguishing two closely space targets. Currently, radio sensing is implemented by radar-like terminals, that work on dedicated bands with \textit{ad-hoc} hardware. However, the integration of an ubiquitous sensing functionality into the existing wireless communication infrastructure cannot underpin the massive deployment of radars. Therefore, integrated sensing and communication (ISAC) systems started gaining interest in recent years~\cite{Liu_survey}. The goal is to design a single system, sharing the whole set of (or part of the) time-frequency-space-hardware resources. Existing literature works on ISAC focused on waveform design trading between communication efficiency and localization accuracy, e.g., see~\cite{Wymeersch2021}.  
%
Stemming from ISAC, the most recent trend is to envision integrated imaging and communication (IIAC) systems. Very few works address IIAC, summarized in \cite{IIAC_lightweight,IIAC_THz_prototyping,FanLiu_imaging,manzoni2024integratedcommunicationimagingdesign}. Work~\cite{IIAC_lightweight} tackles the imaging computational burden by proposing a low-complexity algorithm. The authors of~\cite{IIAC_THz_prototyping} focus on terahertz portable devices, while~\cite{FanLiu_imaging} considers integrating communication and synthetic aperture imaging for low-altitude platform. Finally, the work~\cite{manzoni2024integratedcommunicationimagingdesign} proposes a novel waveform design for IIAC. 

Notwithstanding the raise of IIAC systems, the issue of sensing in non-line-of-sight (NLOS) remains open, limiting the coverage of IIAC networks working above $10$ GHz. NLOS sensing has been explored in various ways, see \cite{9468353,9547412}. All existing works make use of an a-priori knowledge of the surrounding geometry or a properly placed metallic mirror, that constrain the sensing system to leverage a specular reflection only. Alternatively, metasurfaces have been proved to be effective for NLOS exploration~\cite{doi:10.1126/science.1210713}. Metasurfaces are 2D structures whose electromagnetic (EM) properties are engineered to control the wavefield interaction, enabling advanced functionalities (anomalous reflection, focusing, and others). At microwaves, metasurfaces are typically manufactured as 2D arrays of tunable unit cells (meta-atoms), often referred to as \textit{EM skins} (EMSs)~\cite{7109827}. When the tuning is dynamic in time and controllable, we refer to reconfigurable intelligent surfaces (RISs), that have been the focus of a vast literature on RIS-aided localization~\cite{9775078,tagliaferri2023reconfigurable}, with NLOS exploration as a key application~\cite{Buzzi_RISforradar_journal,9511765}. Recently, literature on RISs for sensing started considering the radiative \textit{near-field} effect, namely the operating condition for which the manipulated wavefront is curved across the RIS. In addition to the straightforward increase of the aperture and spatial resolution~\cite{9838638}, new degrees of freedom brought by near-field operation allow exploiting a large EMS/RIS in place of a large active antenna array, reducing cost and energy consumption \cite{9709801}.

Still, very few works investigate the metasurfaces potential for imaging \cite{9299878,torcolacci2023holographic}. The authors of \cite{9299878} address the imaging problem as an inverse scattering reconstruction in far-field, while the authors of \cite{torcolacci2023holographic} tackled the imaging with RISs in both LOS and NLOS scenarios, jointly designing the illumination pattern and the RIS coefficients. RISs have been also considered in the IIAC scenario in \cite{Alkhateeb23_imaging_comm}, where the authors exploit RIS-aided imaging to reduce the communication beam training, by properly designing the RIS phase configuration.

\textbf{Contribution}: This paper proposes a novel IIAC system for NLOS exploration. A BS sweeps an IIAC beam towards a anomalous reflection plane, that is properly configured with a periodic phase pattern in space and time to cover a desired region of interest (ROI) in NLOS, while achieving high-resolution imaging of the same area. Noticeably, the reflection plane changes its configuration in time but it does not necessitates any control; this marks a leap forward compared to RISs, and lowers the implementation costs, allowing much wider \textit{effective apertures}. Moreover, we detail a \textit{modular} design of the plane, that jointly with the BS sweeps can provide \textit{near-field} imaging with successive far-field acquisitions. This work differs from our previous one \cite{bellini2024sensingnlosstroboscopicapproach}, that exploits the motion of the source and the ROI, and it does not consider the time variability of the plane. 
Numerical results prove the benefits of our system with respect to state of the art.

\textbf{Organization and Notation}: The paper is organized as follows: Sect. \ref{sec:strobosens} illustrates the proposed IIAC system, Sect. \ref{sec:system_model} outlines the system model, Sect. \ref{sect:system_design} details the system design, Sect. \ref{sec:ImProcess} shows the results while Sect. \ref{sec:Conclusion} concludes the paper. 
We adopt the following notation: bold lower-case letters stand for column vectors. With $\mathbf{a}\sim\mathcal{CN}(\boldsymbol{\mu},\mathbf{C})$ we denote a circularly complex multi-variate Gaussian random variable with mean $\boldsymbol{\mu}$ and covariance matrix $\mathbf{C}$. $|\mathcal{A}|$ denote the cardinality of set $\mathcal{A}$. $\mathbb{R}$ and $\mathbb{C}$ denote, respectively, the set of real and complex numbers. $\delta_n$ is the Kronecker delta function.

\section{Multi-View IIAC Principle}\label{sec:strobosens}
\begin{figure}[!t]
\centering    \includegraphics[width=0.6\columnwidth]{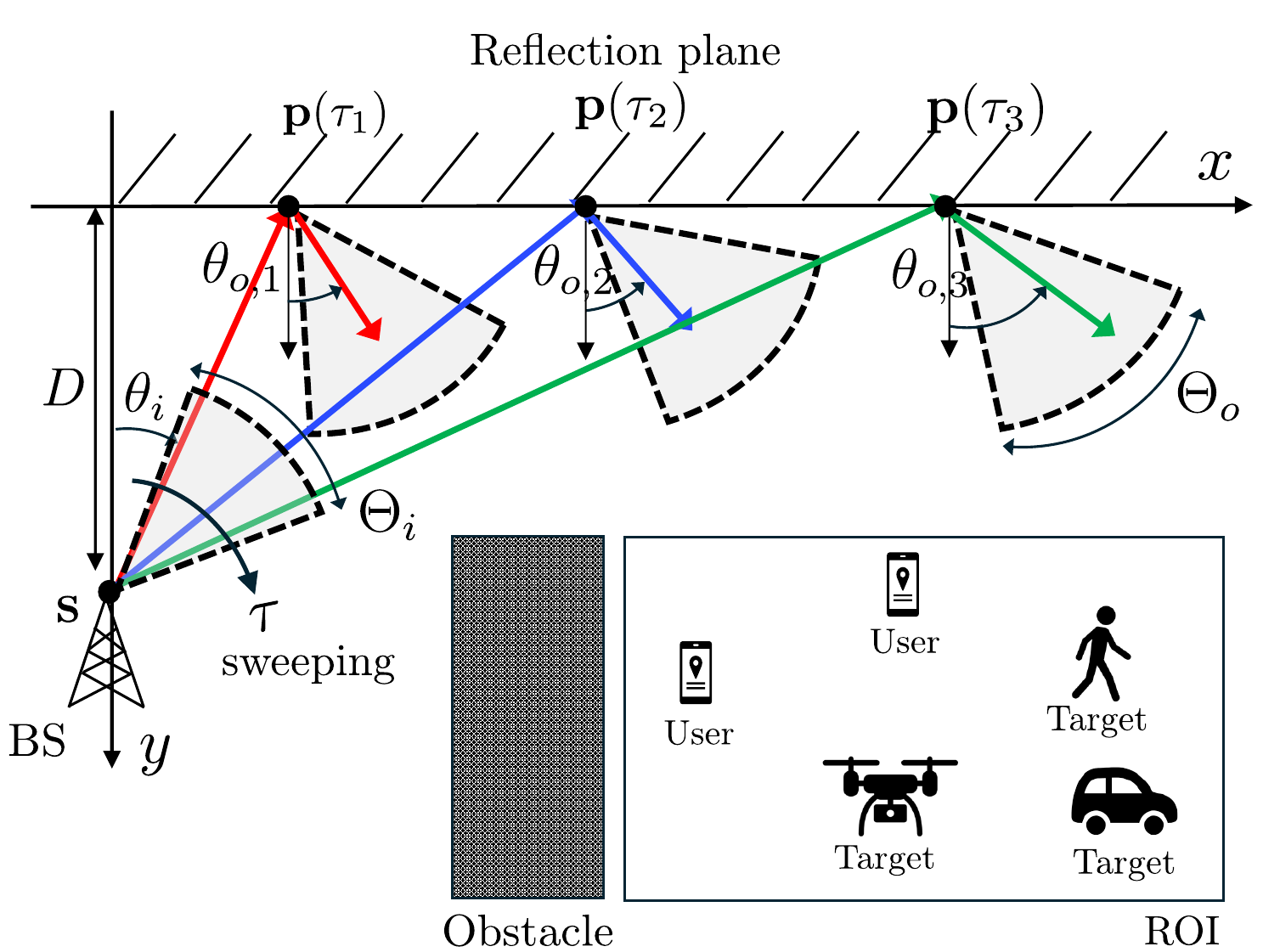}
    \caption{Considered IIAC system: a BS sweeps an IIAC beam towards a reflection plane, that is configured to allow the exploration of a ROI in NLOS, where both users and targets are present. }
    \label{fig:scenario_BS}
\end{figure}

Let us consider the 2D scenario depicted in Fig. \ref{fig:scenario_BS}, where a generic source (e.g., a BS) is located in $\mathbf{s} = \left(0, D\right)$, at distance $D$ from a reflection plane, deployed along $x$. At snapshot time $\tau$ (that we herein refer to \textit{slow-time}), the BS transmits an IIAC signal in direction $\theta_i(\tau)$ via beamforming, changing the pointing angle every $\Delta\tau$ seconds within a discrete codebook $\Theta_i$, the latter spanning an angular interval $\Delta \theta_i$.
The signal impinges a purposely deployed reflection plane in $\mathbf{p}(\tau) = (D\tan \theta_i(\tau),0)$. The plane is capable of providing a space-time varying reflection angle $\theta_o(x,\tau)$ that depends on both the incidence coordinate $x = D\tan \theta_i(\tau)$ and time $\tau$. The dependence on space $x$ maps into a dependence on time $\tau$ due to the BS sweeping $\theta_i(\tau)$, while the explicit dependence on $\tau$ is enforced by the reflection plane at each fixed coordinate $x$. The reflection angle $\theta_o(x,\tau)$ follows a periodic pattern in both space and time, spanning a discrete set of angles $\Theta_o$, of width $\Delta \theta_o$. In such a system, the BS implements a double beam sweeping in $T = |\Theta_i| \Delta \tau$ seconds, serving the users within a desired ROI in NLOS while forming a high-resolution image.

\begin{figure}[!b]
\centering
\subfloat[][]{\includegraphics[width=0.75\columnwidth]{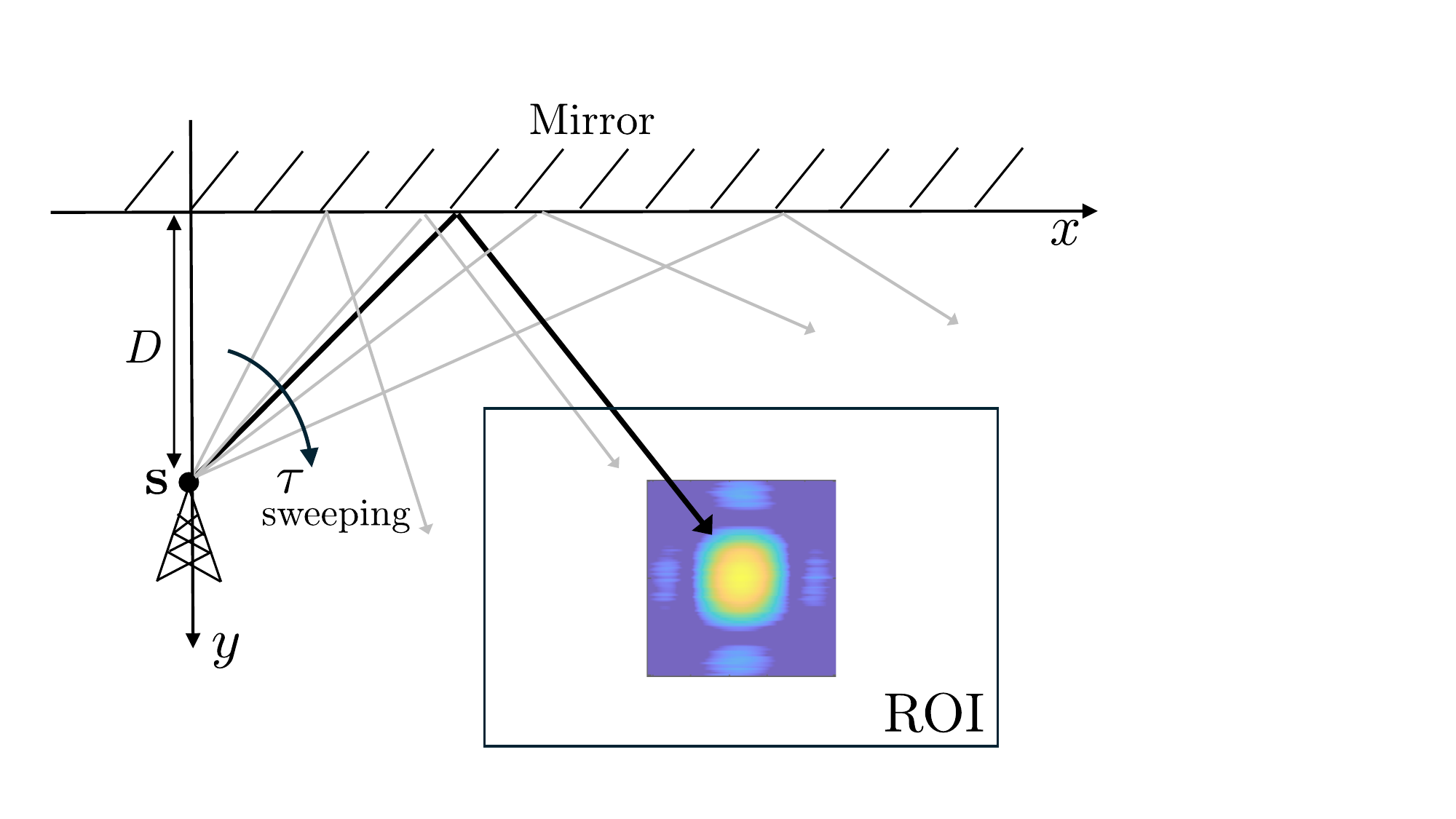}}\\
\subfloat[][]{\includegraphics[width=0.75\columnwidth]{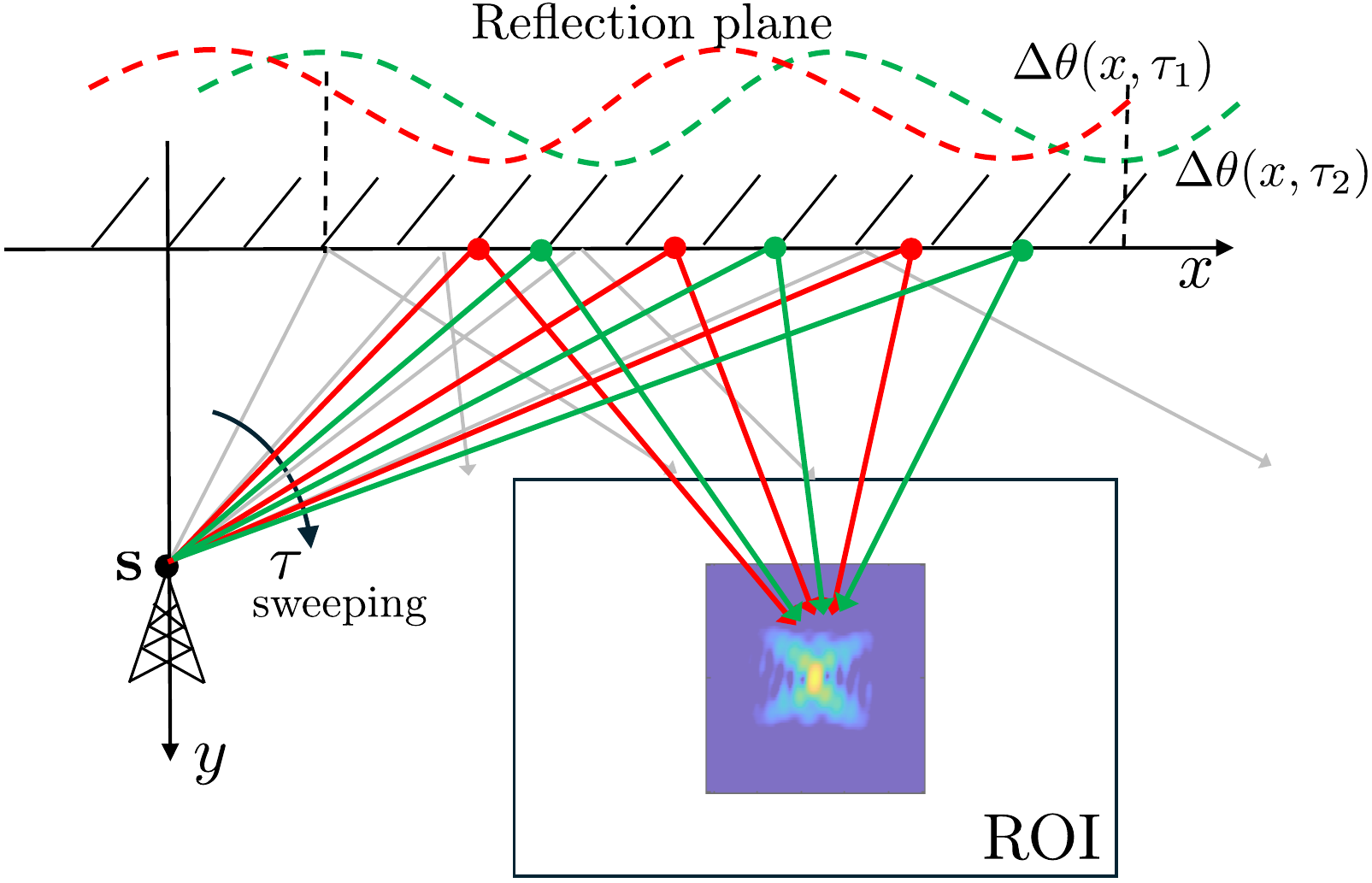}}
    \caption{
    The key idea of multi-view NLOS imaging: by a periodic spatial and temporal configuration of the reflection plane, the BS gathers multiple ($P > 1$) views of the same target, increasing the resolution w.r.t. the use of a bare mirror. Red and green lines refer to different temporal configurations of the plane, providing spatial diversity. }
    \label{fig:enRes}
\end{figure}

In this setup, consider a ROI of size $\Delta_x$ and $\Delta_y$ along $x$ and $y$, centered in $\mathbf{r}^* = (r^*_x, r^*_y)$, and a generic target $\mathbf{r} = (r_x, r_y)$ within the ROI\footnote{A single target is herein instrumental to illustrate the multi-view IIAC system, that is able to image \textit{all} the targets within the ROI.}. The reflection angle at time $\tau$ is:
\begin{equation}\label{eq:theta_o}
    \theta_o(\tau) = \theta_i(\tau) + \Delta\theta(x,\tau)\big\rvert_{x = D \tan \theta_i(\tau)}
\end{equation}
where the reflection plane implements the space-time varying angular difference $\Delta\theta(x,\tau)$. The latter is herein designed as:
\begin{equation}\label{eq:angular_difference}
\begin{split}
    \Delta\theta(x,\tau) = \overline{\theta}_o - \overline{\theta}_i & + \frac{\Delta\theta_{o}}{2}\cos\left(\frac{2\pi}{\Lambda_x} x - \frac{2\pi}{\Lambda_\tau} \tau\right) 
\end{split}
\end{equation}
where $\Lambda_x$ and $\Lambda_\tau$ are the spatial and temporal periodicity, respectively, while $\overline{\theta}_i$ and $\overline{\theta}_o$ denote the central angles of $\Theta_i$ and $\Theta_o$. With such a system, the BS can explore the ROI with a single sweep, serving all the communication users in the ROI within $T$ seconds. In that, the time-varying component of $\Delta\theta(x,\tau)$ is only needed for imaging, as detailed in Section \ref{sect:system_design}.
The problem is to sense the target $\mathbf{r}$. For a given BS position $\mathbf{s}$ and $\theta_i$, the reflection angle $\theta_o$ that guarantees to intercept (illuminate) the target is:
\begin{equation}
    \theta_o\left(\theta_i | \mathbf{s}, \mathbf{r}\right) = \arctan\left(\frac{r_x - D\tan\theta_i}{r_y}\right).
\end{equation}
Given \eqref{eq:angular_difference}, the target will be effectively illuminated only in those time instants obtained by finding the $P$ roots of $\theta_o(\tau) = \theta_o\left(\theta_i | \mathbf{s}, \mathbf{r}\right)$, $\tau \in [0, T]$. If $P >1$, it follows that $\theta_o(\tau_1) \neq \theta_o(\tau_2) \neq ... \neq \theta_o(\tau_P)$, the target is illuminated from \textit{multiple view-points}, and the resolution of the image may increase, as exemplified in Fig. \ref{fig:enRes}. The key observation is that, by multiple BS sweeps and a time-varying $\Delta\theta(x,\tau)$, the set of $P$ roots changes with time, asymptotically covering the entire portion of the reflection plane illuminated by the BS sweeping. The image resolution, therefore, attains its maximum value that only depends on $\Delta\theta_i$ and \textit{not} by the number of BS antennas. 

\textit{Remark}: The image resolution can possibly improve along both $x$ and $y$ w.r.t. BS capabilities, thanks to the near-field effect for large effective apertures. Noticeably, near-field imaging is obtained by successive far-field acquisitions (considering the sweeping of a narrow beam), allows simplifying the design of the system \cite{10541333}.

\textit{Remark 2}: The reflection plane, although time-varying, does not necessitates of a dedicated control, lowering the cost of the implementation.

\section{System Model}\label{sec:system_model}
\begin{figure}[!b]
\centering
    \includegraphics[width=0.75\columnwidth]{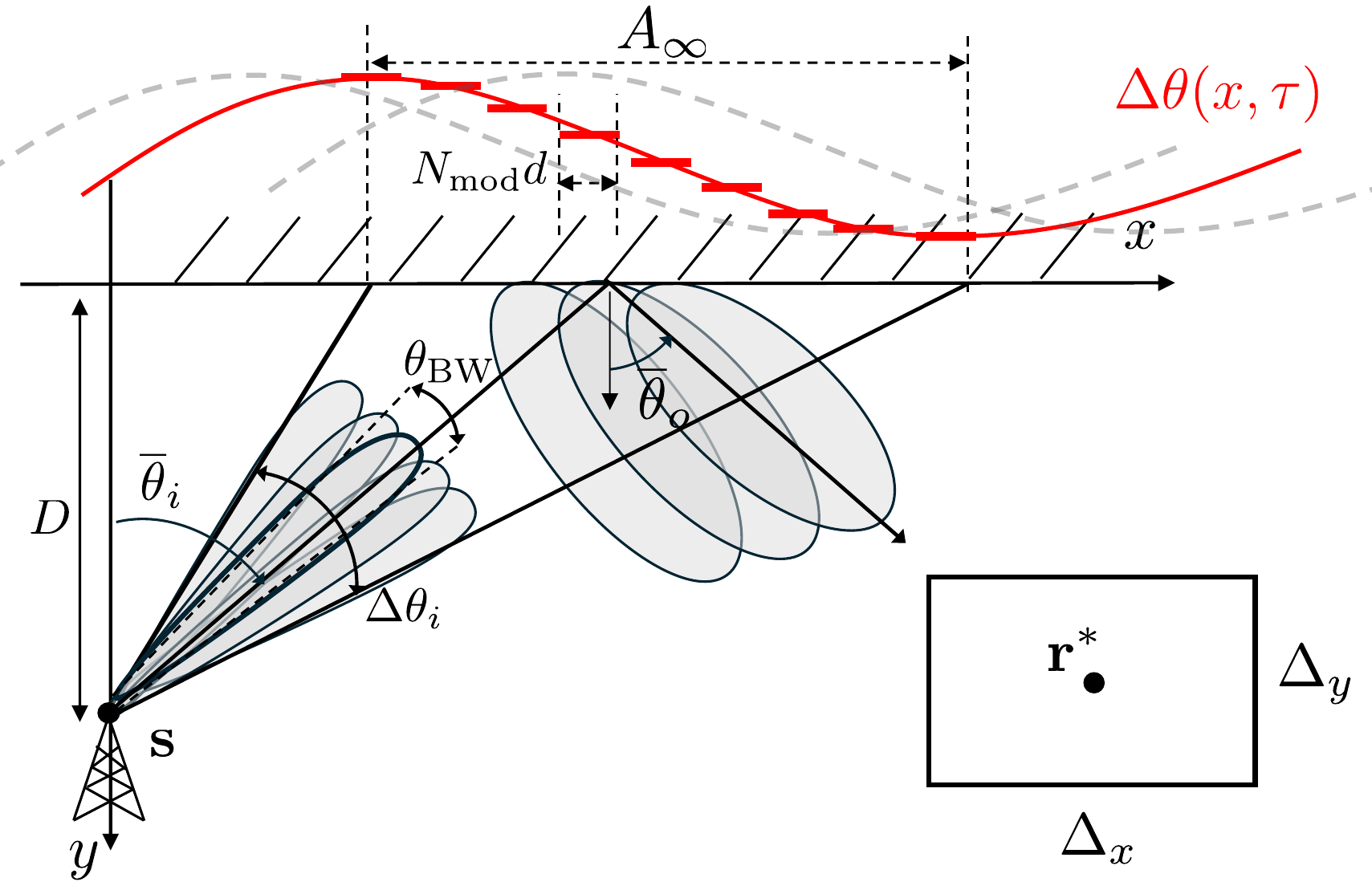}
    \caption{Sketch of the considered geometry. }
    \label{fig:sys_quantities}
\end{figure}

The system model stems from Section \ref{sec:strobosens}. We herein detail, for brevity, the system model for the sensing acquisition only, whereby the communication counterpart follows the standard modeling in \cite{di2020smart}. The IIAC BS operates at frequency $f_0$ and it is equipped with a full-duplex uniform linear array made of $K$ antennas, deployed along $x$ and spaced by $d=\lambda_0/2$ ($\lambda_0$ is the carrier wavelength). The BS aperture is thus $A=Kd$. The BS emits the pass-band IIAC signal: 
\begin{equation}\label{eq:TX_signal}
    g_\text{RF}(t,\tau) = g_\text{BB}(t,\tau) e^{j2\pi f_0 t}
\end{equation}
where $g_\text{BB}(t,\tau)$ is the base-band signal with bandwidth $B$, function of the \textit{fast-time} $t$ and of the slow time $\tau$\footnote{The transmitted signal can be a frame of a orthogonal frequency division multiplexing (OFDM) waveform, whereby each symbol is beamformed in a different direction $\theta_i(\tau)$.}. BS employs a phase-only beamforming vector $[\mathbf{f}(\tau)]_k = e^{j \frac{2 \pi}{\lambda_0} d k \sin \theta_i(\tau)}$ to focus the emitted signal $g_\text{BB}(t,\tau)$ towards direction $\theta_i(\tau)$, over a beamwidth $\theta_\text{BW}(\tau) \simeq \lambda_0/(A \cos \theta_{i}(\tau))$.
%
%
The reflection plane is made by a planar metasurface, composed by $N$ equally spaced elements along the $x$ axis\footnote{The 2D system model outlined in this paper is instrumental to describe the IIAC imaging system design on the azimuth plane. In practice, planar metasurfaces deployed on the $xz$ plane are required to guarantee a sufficient signal-to-noise ratio (SNR). The 3D modeling follows from the present with due adaptations.}, whose $n$th element is located in $\mathbf{p}_n = (x_0 + nd, 0), n = -\frac{N}{2}+1,...,\frac{N}{2}$, where $x_0$ is the deployment offset w.r.t. the origin of the axes. 
At time instant $\tau$, the BS illuminates a portion of metasurface made by
\begin{equation}
    M(\tau) \approx \frac{D \, \theta_\text{BW}(\tau)}{d \sin \theta_{i}(\tau) \cos \theta_{i}(\tau) } \ll N
\end{equation}
elements within the meta-atoms set $\mathcal{M}(\tau)$. 

The general model of the received signal at the BS is reported in \eqref{eq:Rx_sig_gen}, due to a target in $\mathbf{r}=(r_x,r_y)$. 
\begin{figure*}[!t]
\begin{equation}
\label{eq:Rx_sig_gen}
\begin{split}
        y(t,\tau) & = \mathbf{h}^H_i(\tau) \mathbf{\Phi}^H(\tau) \mathbf{h}_o(\tau) \, \Gamma \, \mathbf{h}_o^H(\tau) \mathbf{\Phi}(\tau) \mathbf{h}_i(\tau) \; g\left(t- \frac{2\left[D_{i}(\tau) \hspace{-0.1cm}+\hspace{-0.1cm} D_{o}(\tau)\right]}{c},\tau\right) + w(t,\tau) \\
        & = \alpha\, e^{-j \frac{4 \pi}{\lambda_0} \left[D_{i}(\tau) + D_{o}(\tau)\right]} \hspace{-0.5cm}\sum_{m,m'\in \mathcal{M}(\tau)} \hspace{-0.4cm} e^{j \phi_m(\tau)}e^{j\phi_{m'(\tau)}}  e^{-j \frac{2\pi d}{\lambda_0}(m+m') \left[ \sin\theta_{i}(\tau) -  \sin\theta_{o}(\tau)\right]} g\left(t- \frac{2\left[D_{i}(\tau) \hspace{-0.1cm}+ \hspace{-0.1cm}D_{o}(\tau)\right]}{c},\tau\right) + w(t,\tau)
\end{split}
\end{equation}
\hrulefill
\end{figure*}
In \eqref{eq:Rx_sig_gen}, $g(t)$ is the base-band signal after matched filter, while
\begin{align}
    [\mathbf{h}_i(\tau)]_{m\in \mathcal{M}(\tau)} &= \frac{\lambda_0}{4 \pi D_i(\tau)} e^{-j \frac{2 \pi}{\lambda_0} [D_i(\tau) + m d \sin \theta_i(\tau)]} \\
    [\mathbf{h}_o(\tau)]_{m\in \mathcal{M}(\tau)} & = \frac{\lambda_0}{4 \pi D_o(\tau)} e^{-j \frac{2 \pi}{\lambda_0} [D_o(\tau) + m d \sin \theta_o(\tau)]}
\end{align}
denote the incident and reflection channel to/from the metasurface, respectively, for the set of illuminated elements $m\in \mathcal{M}(\tau)$. $D_i(\tau) = \| \mathbf{p}(\tau)-\mathbf{s}\|$ and $D_o(\tau) = \| \mathbf{r} - \mathbf{p}(\tau)\|$ are the BS-metasurface and metasurface-target distances at time $\tau$, respectively (considering the beam center). Time-varying diagonal matrix $\mathbf{\Phi}(\tau)\in\mathbb{C}^{M \times M}$ is the phase-configuration of the metasurface, whose $m$th diagonal element is $e^{j\phi_m(\tau)}$, to be designed according to \eqref{eq:angular_difference}. Term $\Gamma$ is the target's reflectivity, while $w(t,\tau)$ is the noise corrupting the received signal with power $\sigma_w^2$. The received signal model \eqref{eq:Rx_sig_gen} holds under both \textit{far-field} and \textit{spatial narrow-band} assumptions, as the first condition implies that the beamwidth $\theta_\text{BW}$ is narrow enough to have a planar wavefront across the illuminated meta-atom sets $\mathcal{M}(\tau)$, while the spatial narrowband condition holds for $1/B \gg M(\tau) d/c \max\{\sin\theta_{i}(\tau), \sin\theta_{o}(\tau)\}$, meaning that the residual propagation delay across the effective array is much less than the pulse duration $1/B$. These assumptions simplify the system model and the derivations, but do not limit the generality of this work.

The union of sets $\mathcal{M}(\tau)$ covered by BS sweeping, i.e., $ \cup_{\substack{\tau}} \mathcal{M}(\tau), \tau \in [0,T]$ gives the \textit{asymptotic aperture}, namely the maximum attainable aperture that maximizes the image resolution within the ROI. The asymptotic aperture is approximated for narrow beamwidth $\theta_\text{BW}$ as
\begin{equation}\label{eq:Aeff}
    A_\infty \simeq D \left[ \tan\left(\overline{\theta}_i \hspace{-0.05cm} + \hspace{-0.05cm} \frac{\Delta \theta_{i}}{2}\right) - \tan\left(\overline{\theta}_i \hspace{-0.05cm}-\hspace{-0.05cm} \frac{\Delta \theta_{i}}{2}\right)\right]
\end{equation}
and it coincides with the total portion of the metasurface illuminated by the BS. In general, it is $A_\infty \gg A$. To obtain an image with the resolution dictated by $A_\infty$, however, it is necessary to enforce a temporal pattern across the metasurface to vary the observation angle towards the ROI, as explained in the following. 

\section{System Design}\label{sect:system_design}


The design of the system has the ultimate goal of enabling high-resolution imaging of a desired ROI while serving the users therein. As detailed in Sect. \ref{sec:strobosens}, by sweeping the transmitted beam of the source and introducing a space-time variation of the reflected beam orientation from the reflection plane, we can illuminate a target within the ROI from multiple viewpoints. For sake of reasoning, the reflection plane is implemented by a sequence of passive EMS (\textit{modules}) to replace $\Delta\theta(x,\tau)$ with a finite set of reflection angles:
\begin{equation}
\label{eq:quant_func}
    \Delta\theta_{\text{mod}}(x,\tau) \hspace{-0.05cm}=\hspace{-0.05cm} \mathcal{Q}_{\Theta_o} \left\{ \overline{\theta}_o \hspace{-0.1cm}-\hspace{-0.1cm} \overline{\theta}_i \hspace{-0.05cm}+\hspace{-0.05cm} \frac{\Delta\theta_{o}}{2}\cos\left(\frac{2\pi}{\Lambda_x} x \hspace{-0.1cm}- \hspace{-0.1cm}\frac{2\pi}{\Lambda_\tau} \tau\right)\right\}
\end{equation}
where $\mathcal{Q}_{\Theta_o}(\cdot)$ is the quantization function over set $\Theta_o$.
The average number of elements per module is therefore $N_{\text{mod}} \simeq \Lambda_x / 2 d |\Theta_o|$.
The required phase gradient to apply at the EMS to implement the desired angular difference is:
\begin{equation}
    \phi(x,\tau) = \frac{2\pi}{\lambda_0} [\sin\left(\overline{\theta}_i\right) - \sin\left(\overline{\theta}_i+ \Delta\theta_{\text{mod}}(x,\tau)\right)] x
\end{equation}
In the following, we detail the designing criteria for the proposed system.

\subsection{Selection of BS Codebook $\Theta_i$}

The codebook set $\Theta_i$ implemented by the BS is 
\begin{equation}
    \Theta_i = \left\{\overline{\theta}_i -\frac{\Delta \theta_{i}}{2} : \delta \theta_i : \overline{\theta}_i + \frac{\Delta \theta_{i}}{2}\right\}
\end{equation}
where $\delta\theta_i$ is the angular sampling. For practical applications where the size of the metasurface $L=Nd$ is limited, $\overline{\theta}_i$ is selected to point to the center of the metasurface, i.e., $\overline{\theta}_i = \arctan\left(x_0/D\right)$. The spanned angular interval $\Delta \theta_{i}$, instead, shall guarantee the illumination of the entire metasurface size, $A_\infty \leq L$. Differently, the sampling interval $\delta\theta_i$ is dictated by the need of avoiding spatial ambiguities within the image of the ROI. To this end, we analyze the derivative of the propagation phase $\varphi(\theta_{i}(\tau) | \mathbf{r}) = \frac{2 \pi}{\lambda_0} [D_{i}(\theta_i) + D_{o}(\theta_i;\mathbf{r})]$ w.r.t. an infinitesimal variation of $\theta_i$, for a given target $\mathbf{r}$ within the ROI. The result is:
\begin{equation}\label{eq:ph_derivative}
\begin{split}
      \frac{\mathrm{d}\varphi (\theta_i|\mathbf{r})}{\mathrm{d}\theta_i } & = \frac{\mathrm{d}}{\mathrm{d} \theta_i} \hspace{-0.1cm} \left[ \frac{4 \pi}{\lambda_0} (D_i(\theta_i) + D_o(\theta_i|\mathbf{r}))\right] \\ 
      & = \frac{4 \pi D}{\lambda_0 \cos^2\theta_i} \left[\sin\theta_i\hspace{-0.1cm} -\hspace{-0.1cm}\frac{r_x \hspace{-0.1cm}- \hspace{-0.1cm}D \tan \theta_i}{\sqrt{r_y^2 + (r_x \hspace{-0.1cm}-\hspace{-0.1cm} D \tan \theta_i)^2}}\right]
\end{split}
\end{equation}
The sampling interval is therefore upper-bounded by the maximum difference between the phase derivatives within $\Delta \theta_{i}$:
\begin{equation}\label{eq:deltathetai}
    \delta \theta_i \leq \frac{ \pi}{\bigg \lvert \underset{\mathbf{r}}{\max} \left\{ \frac{\mathrm{d}\varphi (\theta_i|\mathbf{r})}{\mathrm{d}\theta_i } \big \rvert_{\theta_i=\frac{\Delta \theta_{i}}{2}}\right\} - \underset{\mathbf{r}}{\min} \left\{  \frac{\mathrm{d}\varphi (\theta_i|\mathbf{r})}{\mathrm{d}\theta_i } \big \rvert_{\theta_i=-\frac{\Delta \theta_{i}}{2}}\right\} \bigg \rvert}
\end{equation}
More insights can be found in \cite{bellini2024sensingnlosstroboscopicapproach}.
%
%
%

\subsection{Selection of Reflection Codebook $\Theta_o$}

The reflection codebook implemented by the EMS is:
\begin{equation}
\label{eq:reflection_codebook}
   \Theta_o = \left\{ \overline{\theta}_o - \frac{\Delta \theta_{o}}{2} : \delta \theta_o :  \overline{\theta}_o + \frac{\Delta\theta_{o}}{2}\right\}
\end{equation}
where $\delta\theta_o$ is the angular sampling in reflection. Angle $\overline{\theta}_o$ depends on the relative position of the ROI w.r.t. the metasurface, i.e., $\overline{\theta}_o = \arctan([r_x^* - x_0]/r^*_y)$. The angular interval $\Delta \theta_{o}$ shall be designed to enable the illumination of the ROI \cite{bellini2024sensingnlosstroboscopicapproach}. The angular sampling $\delta\theta_o$, and consequently the number of EMS modules $|\Theta_o|$, is chosen to ensure that each target in the ROI is illuminated by at least one (and possibly more) EMS modules. The condition is set by imposing that the reflected beams by adjacent modules are partially overlapped, yielding: 
\begin{equation}\label{eq:deltathetao}
   \delta \theta_o \leq  \frac{1}{2} \underset{\theta_{o} \in \Theta_o}{\min}\left\{\frac{\lambda_0}{N_\text{mod} d \cos\theta_{o}} \right\} 
\end{equation}
where we select the minimum beamwidth over $\Theta_o$. Notice that \eqref{eq:deltathetao} implies a constraint on the maximum allowed size of each module.

\subsection{Selection of Spatial and Temporal Periodicity $\Lambda_x$ and $\Lambda_\tau$}

The optimal choice of the spatial component of $\Delta \theta(x,\tau)$ is trade-off between image resolution and sidelobe ratio, and it can be demonstrated that a reasonable choice is $\Lambda_x = 2 A_\infty$, namely the spatial period shall be double the illuminated area by the BS \cite{bellini2024sensingnlosstroboscopicapproach}. 

The role of the temporal periodicity $\Lambda_\tau$, instead, is to enable a high-quality sensing within the ROI exploiting multiple BS sweeps and different metasurface configurations. For a fixed temporal configuration of the metasurface (i.e., for $\Lambda_\tau\rightarrow \infty$), the BS sweeps exploits a space-varying angular difference $\Delta \theta(x)$. In this latter configuration, each target within the ROI is, in general, illuminated by a different set of observation points across the metasurface (see Fig. \ref{fig:enRes}b), leading to a space-varying image quality across the ROI (\textit{target unfairness}). If this might be acceptable for communication, where the goal is to have the coverage of each single point in the ROI, imaging shall attain a space-invariant spatial resolution and sidelobe level. To this aim, if the the modules of the metasurface change their phase configuration in time, over multiple BS sweeps $\Delta \theta(x,\tau)$, the image quality in \textit{each} location within the ROI reaches its maximum, since the target is observed by multiple (always different) set of reflection points on the plane. Therefore, a reasonable choice is to select $\Lambda_\tau \gg T$, in practice $\Lambda_\tau \simeq |\Theta_o|T$. The final image is obtained after $|\Theta_o|$ sweeps, at the single snapshot each EMS module explores an angle in~$\Theta_o$. 

\begin{figure*}[!t]
    \centering
    \subfloat[][Mirror \cite{9468353,9547412}]{\includegraphics[width=0.29\columnwidth]{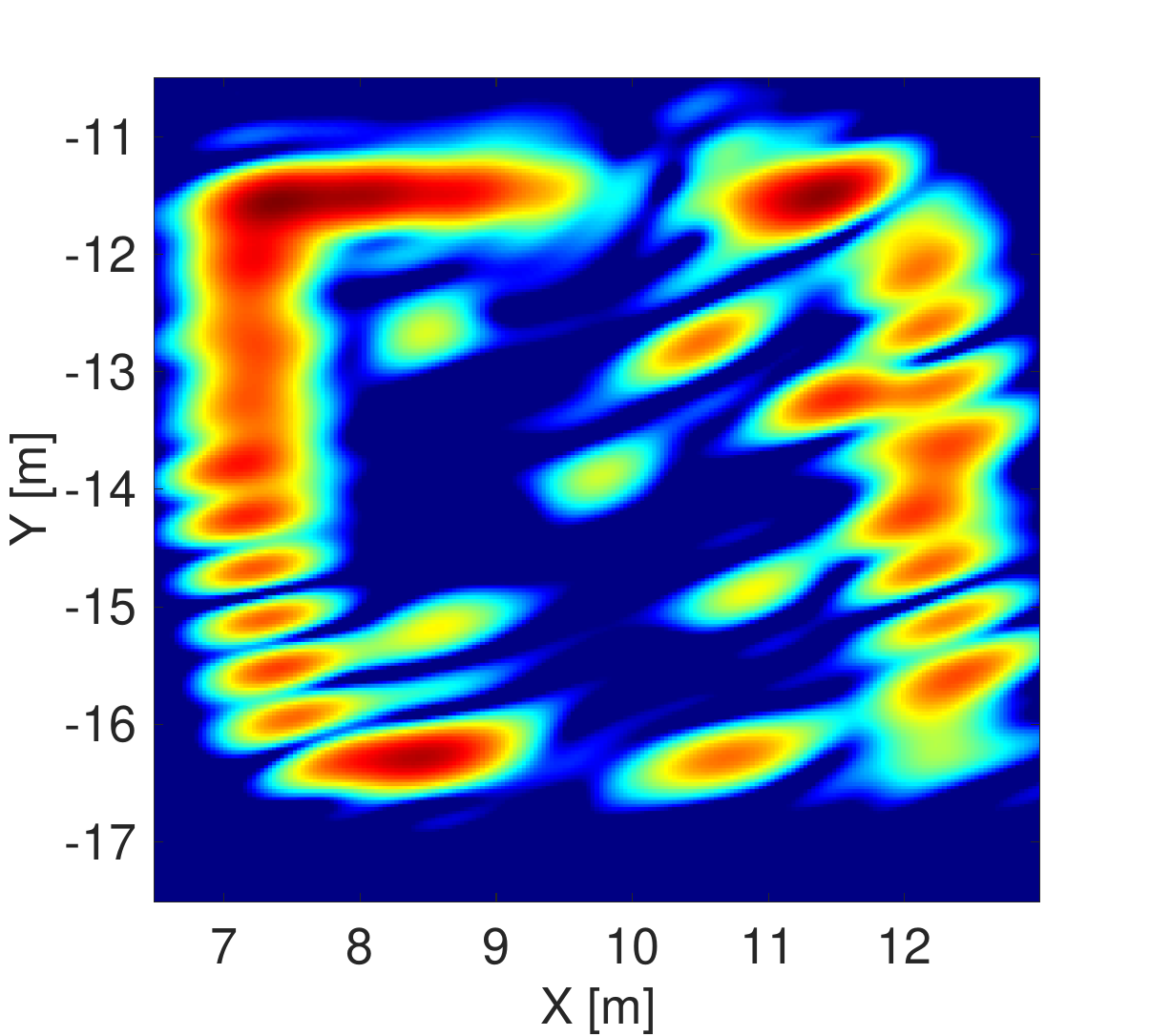}\label{subfig:mirror}} 
    \subfloat[][Single-sweep, multi-view]{\includegraphics[width=0.29\columnwidth]{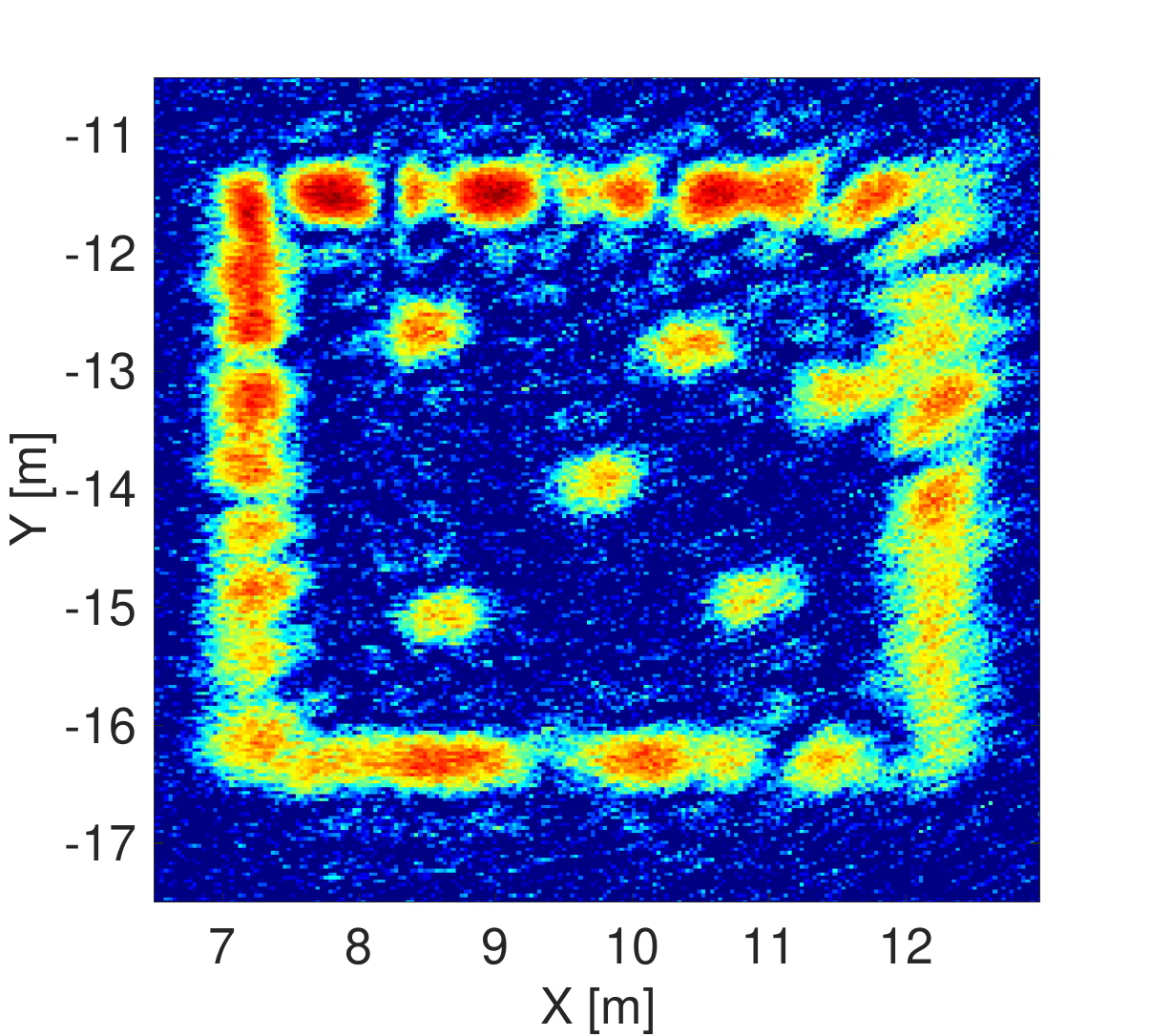}\label{subfig:single_sweep}} 
    \subfloat[][Multi-sweep, multi-view]{\includegraphics[width=0.29\columnwidth]{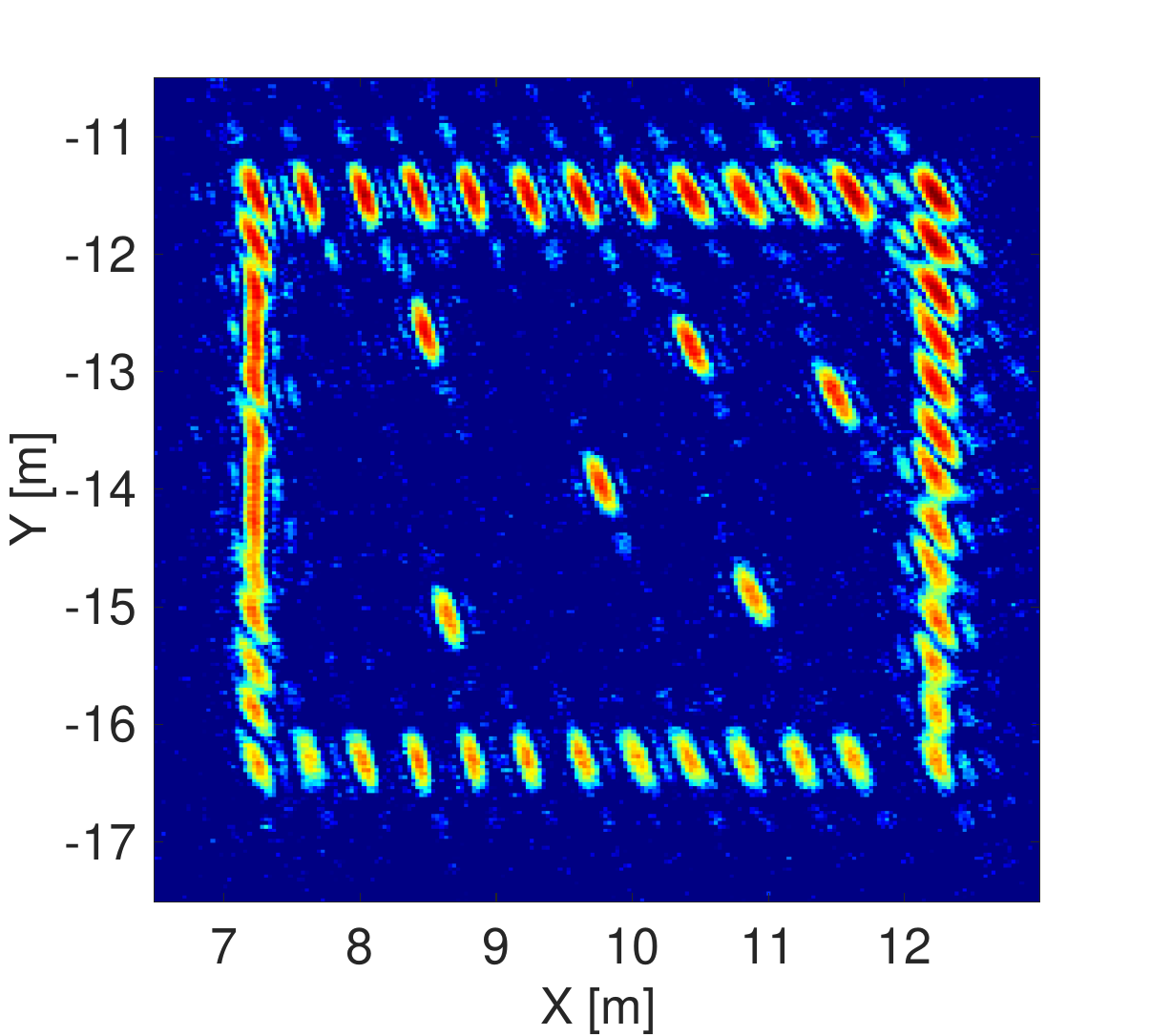}\label{subfig:multi_sweep}}
    \caption{Image of a complex target shape obtained with (a) a metallic mirror \cite{9468353,9547412} and (b) the proposed multi-view system, using a single sweep, i.e., with no reconfiguration of the plane. The image in (c) is obtained with by combining multiple sweeps over $\Theta_i$, with different temporal configurations of the plane. }
    \label{fig:imaging}
\end{figure*}

\section{Image Synthesis and Results}
\label{sec:ImProcess}

The image of the ROI is synthesized by back-projection (BP) in time. BP is the matched filtering over the fast and slow time, as it consists in a correlation between the received signal and the noiseless model signal that would be received by back-scattering from a target in $\mathbf{x} = (x,y)$ within the grid representing the ROI. The complex image of a ROI yields:
\begin{equation}\label{eq:image_BP}
    I(\mathbf{x}) {\simeq} \sum_{\tau} y\left(t = \frac{2 [D_{i}(\tau) + D_{o}(\tau;\mathbf{x})]}{c}, \tau\right) e^{j \varphi(\theta_{i}(\tau) | \mathbf{x})}
\end{equation}
coherently combining the received echoes for each snapshot time angle $\tau \in [0, |\Theta_o|T]$. Here $D_{o,\ell}(\mathbf{x}) = \| \mathbf{x} - \mathbf{p}(\tau)\|$ and $\varphi(\tau | \mathbf{x}) = \frac{2 \pi}{\lambda_0}[D_{i}(\tau) + D_{o}(\tau;\mathbf{x})]$ is the propagation phase. 

To assess the performance of our proposed system, we herein discuss few imaging examples in terms of resolution quality. The simulation parameters are as follows: carrier frequency $f_0 = 28$ GHz, employed bandwidth $B = 400 $ MHz. The BS is located at distance $D = 5$ m from the reflection plane, emitting the signal over a beamwidth of $\theta_{\text{BW}}\simeq 2.5$ deg, sweeping within a codebook $\Theta_i$ of size $\Delta\theta_{i} = 10$ deg centered around $\overline{\theta}_i = 30$ deg. The reflection plane is made by a sequence of adjacent modules, whose horizontal size $N_{\text{mod}}d = 20$ cm is set according to Sect. \ref{sect:system_design}, as the space-time varying reflection angle pattern \eqref{eq:quant_func} spans a codebook $\Theta_o$ with cardinality $|\Theta_o| =15$ over a spatial periodicity $\Lambda_x = 6$ m. The ROI of size $\Delta_x = \Delta_y = 5$ m is located in $\mathbf{r}^* = (9.5, -14)$ m. The temporal periodicity is $\Lambda_\tau = 150$ ms. The useful signal at received side is corrupted by thermal noise with power $\sigma_w^2=-87$ dBm.

Fig. \ref{fig:imaging} shows an example of radar imaging for a target of complex shape, composed of a square and isolated targets within. For simplicity, the blockage effect was not considered. The proposed multi-view system, whose results are reported in Figs. \ref{subfig:single_sweep} and \ref{subfig:multi_sweep}, is compared to a mirror-based imaging system from \cite{9468353,9547412} (Fig. \ref{subfig:mirror}). 
The images are all normalized to their maximum values and blockage effect is not considered to showcase the benefits of our proposal. Fig. \ref{subfig:single_sweep} considers a single BS sweep, thus a fixed spatial configuration of the metasurface plane, while Fig. \ref{subfig:multi_sweep} is obtained by coherent summation of successive sweeps. In case of a mirror, the achievable image resolution is dictated by the BS aperture (thus by beamwdith $\theta_{\text{BW}}$) as well as bandwidth $B$. Differently, our system attains a better result by exploiting the aperture provided by the larger metasurface plane. However, a single BS sweep and metasurface configuration is not sufficient to achieve $A_\infty$. The reason is that with $\Delta\theta(x,\tau)=\Delta\theta(x)$, each target is is only observed with a limited spatial diversity, corresponding to $P$ selected points (see Fig. \ref{fig:enRes}b). By multiple sweeps and $\Delta\theta(x,\tau)$, instead, targets are asymptotically illuminated by \textit{all} the metasurface, maximizing the quality of the image in terms of resolution and SNR (Fig. \ref{subfig:multi_sweep}).
\begin{figure}[!t]
    \centering
    \subfloat[][$r_y = 25$ m]{\includegraphics[width=0.44\columnwidth]{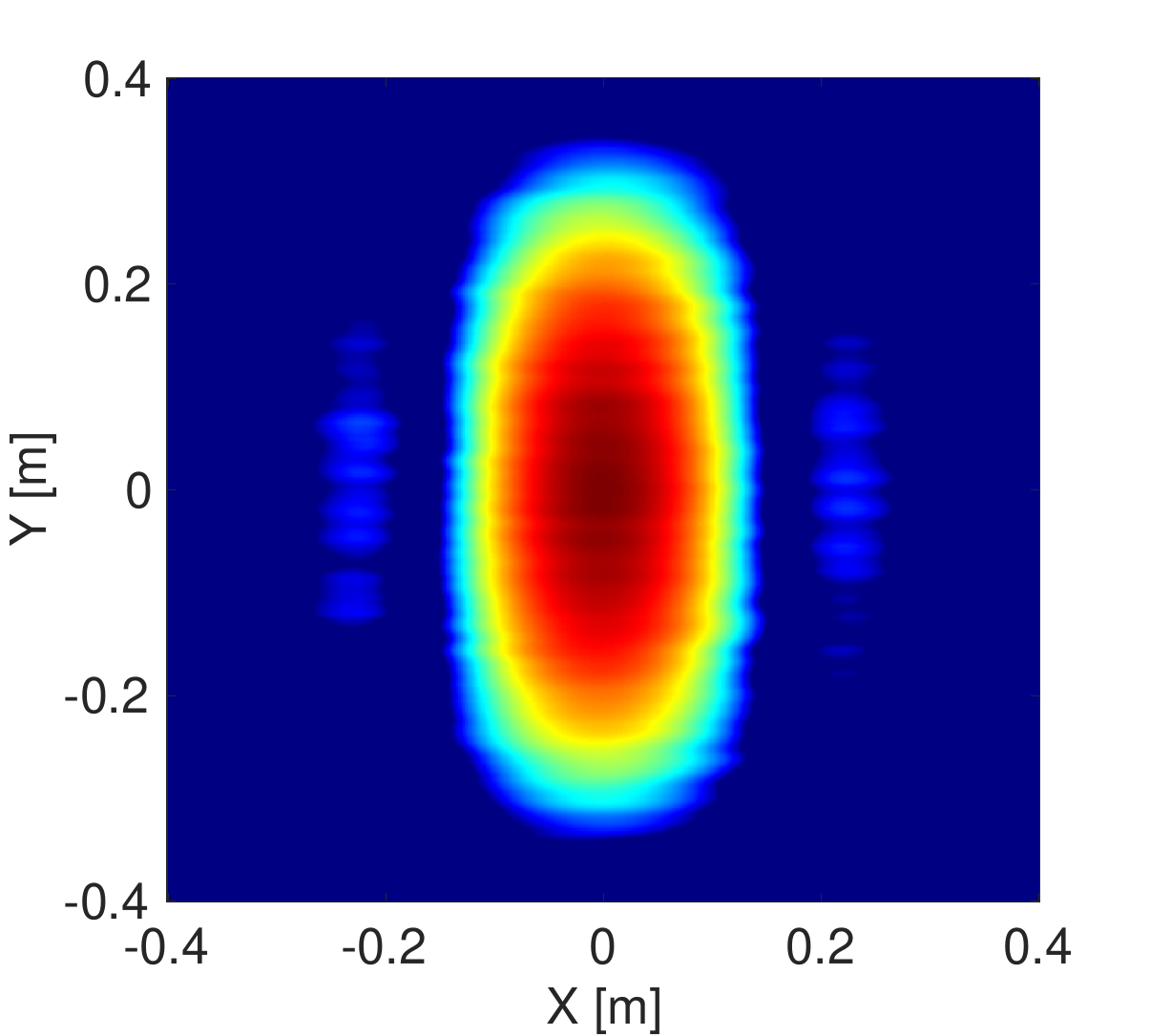}\label{subfig:25m}} 
    \subfloat[][$r_y = 2$ m]{\includegraphics[width=0.44\columnwidth]{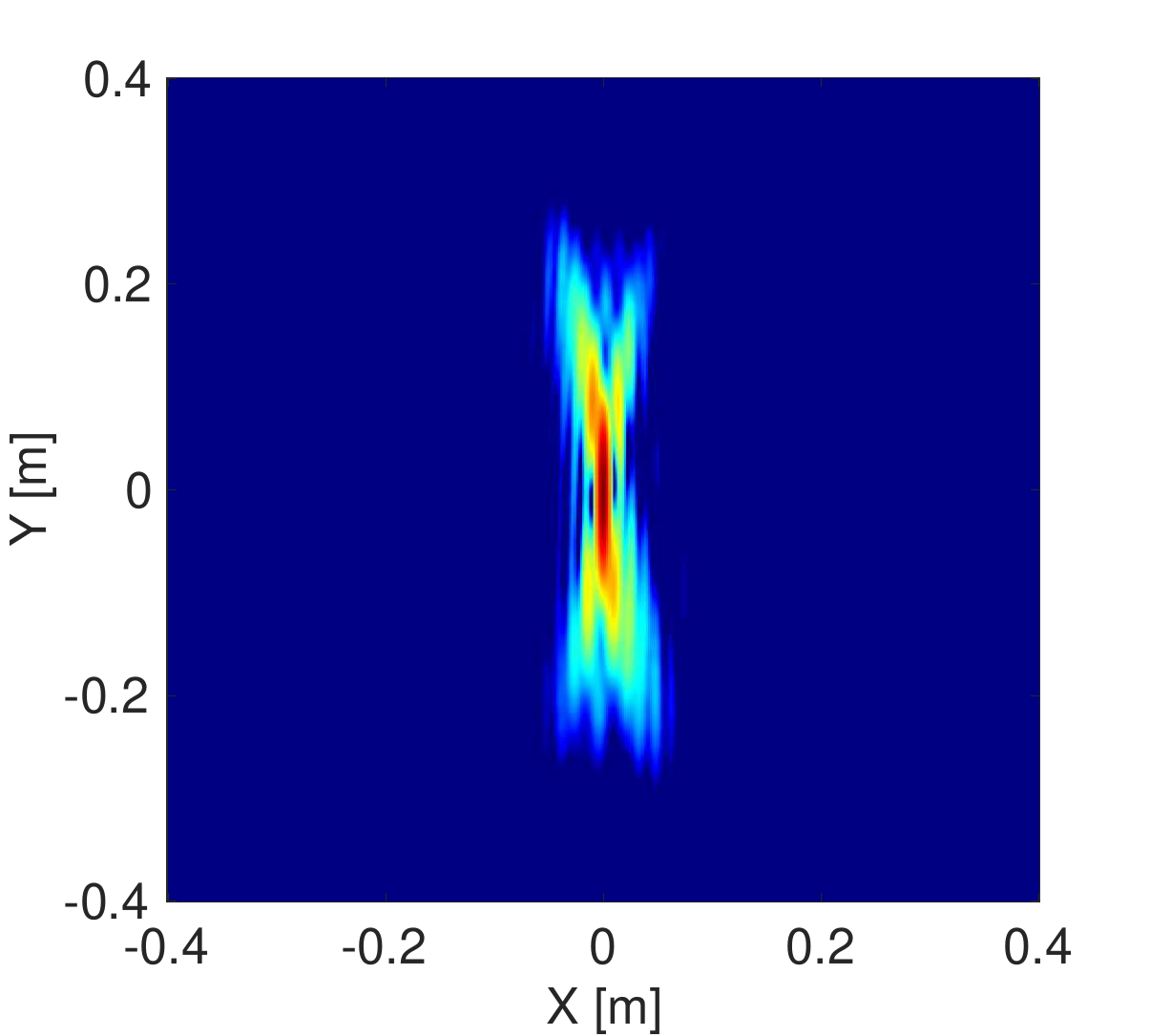}\label{subfig:2m}}
    \caption{Image of a point target for varying distance $r_y$ from the reflection plane: (a) 25 meters and (b) 2 meters. The near-field effect allows gaining range resolution (along $y$) thanks to multiple diverse observation directions, whereas the system is fully designed under far-field assumption.}
    \label{fig:near-far-field}
\end{figure}

The last result in Fig. \ref{fig:near-far-field} shows the image of a point target varying the distance with the reflection plane ($r_y$). The result is obtained for $\overline{\theta}_o = 0$ deg, such that to evaluate the enhancement of the range resolution due to near-field effect. Provided that a sufficiently large portion of metasurface is effectively illuminated through multiple sweeps, the multi-view system allows improving not only the cross-range resolution along $x$, but also the range resolution along $y$. Noticeably, in both cases (Figs. \ref{subfig:25m} and \ref{subfig:2m}) the system design is in far-field. Near-field is obtained from the coherent combination of successive measurements.

\section{Conclusions}
\label{sec:Conclusion}

This work proposes and discusses a novel IIAC system for imaging in NLOS, aided by a properly configured metasurface whose intent is to illuminate a desired ROI. 
The system leverages on a contiguous sweep of the source beam over the reflection plane, which implements a space-time varying angular reflection function to cover the ROI. 
The metasurface consists in a sequence of passive EMS, whose joint far-field configuration allows increasing the resolution of the ROI image w.r.t. BS capabilities.
The presented imaging results show the potential benefits of our method with respect to existing techniques.

\bibliographystyle{IEEEtran}
\bibliography{bibtex/bib/Bibliography,bibtex/bib/Bibliography_TWC}

\begin{thebibliography}{10}
\providecommand{\url}[1]{#1}
\csname url@samestyle\endcsname
\providecommand{\newblock}{\relax}
\providecommand{\bibinfo}[2]{#2}
\providecommand{\BIBentrySTDinterwordspacing}{\spaceskip=0pt\relax}
\providecommand{\BIBentryALTinterwordstretchfactor}{4}
\providecommand{\BIBentryALTinterwordspacing}{\spaceskip=\fontdimen2\font plus
\BIBentryALTinterwordstretchfactor\fontdimen3\font minus \fontdimen4\font\relax}
\providecommand{\BIBforeignlanguage}[2]{{%
\expandafter\ifx\csname l@#1\endcsname\relax
\typeout{** WARNING: IEEEtran.bst: No hyphenation pattern has been}%
\typeout{** loaded for the language `#1'. Using the pattern for}%
\typeout{** the default language instead.}%
\else
\language=\csname l@#1\endcsname
\fi
#2}}
\providecommand{\BIBdecl}{\relax}
\BIBdecl

\bibitem{Wymeersch6G_ISAC}
N.~González-Prelcic, M.~F. Keskin, O.~Kaltiokallio, M.~Valkama, D.~Dardari, X.~Shen, Y.~Shen, M.~Bayraktar, and H.~Wymeersch, ``The integrated sensing and communication revolution for 6g: Vision, techniques, and applications,'' \emph{Proceedings of the IEEE}, pp. 1--0, 2024.

\bibitem{10287134}
S.~E. Trevlakis, A.-A.~A. Boulogeorgos, D.~Pliatsios, J.~Querol, K.~Ntontin, P.~Sarigiannidis, S.~Chatzinotas, and M.~D. Renzo, ``Localization as a key enabler of 6g wireless systems: A comprehensive survey and an outlook,'' \emph{IEEE Open Journal of the Communications Society}, pp. 1--1, 2023.

\bibitem{tagliaferri2023cooperative}
D.~Tagliaferri, M.~Manzoni, M.~Mizmizi, S.~Tebaldini, A.~V. Monti-Guarnieri, C.~M. Prati, and U.~Spagnolini, ``Cooperative coherent multistatic imaging and phase synchronization in networked sensing,'' \emph{IEEE Journal on Selected Areas in Communications}, pp. 1--1, 2024.

\bibitem{Chetty2022_CRB}
A.~Liu, Z.~Huang, M.~Li, Y.~Wan, W.~Li, T.~X. Han, C.~Liu, R.~Du, D.~K.~P. Tan, J.~Lu, Y.~Shen, F.~Colone, and K.~Chetty, ``A survey on fundamental limits of integrated sensing and communication,'' \emph{IEEE Communications Surveys \& Tutorials}, vol.~24, no.~2, pp. 994--1034, 2022.

\bibitem{Liu_survey}
F.~Liu, Y.~Cui, C.~Masouros, J.~Xu, T.~X. Han, Y.~C. Eldar, and S.~Buzzi, ``Integrated sensing and communications: Toward dual-functional wireless networks for {6G} and beyond,'' \emph{IEEE Journal on Selected Areas in Communications}, vol.~40, no.~6, pp. 1728--1767, 2022.

\bibitem{Wymeersch2021}
M.~F. Keskin, V.~Koivunen, and H.~Wymeersch, ``Limited feedforward waveform design for ofdm dual-functional radar-communications,'' \emph{IEEE Transactions on Signal Processing}, vol.~69, pp. 2955--2970, 2021.

\bibitem{IIAC_lightweight}
X.~Li and Y.~Chen, ``Lightweight 2d imaging for integrated imaging and communication applications,'' \emph{IEEE Signal Processing Letters}, vol.~28, pp. 528--532, 2021.

\bibitem{IIAC_THz_prototyping}
O.~Li, J.~He, K.~Zeng, Z.~Yu, X.~Du, Y.~Liang, G.~Wang, Y.~Chen, P.~Zhu, W.~Tong, D.~Lister, and L.~Ibbotson, ``Integrated sensing and communication in 6g a prototype of high resolution thz sensing on portable device,'' in \emph{2021 Joint European Conference on Networks and Communications \& 6G Summit (EuCNC/6G Summit)}, 2021, pp. 544--549.

\bibitem{FanLiu_imaging}
B.~Zheng and F.~Liu, ``Random signal design for joint communication and sar imaging towards low-altitude economy,'' \emph{IEEE Wireless Communications Letters}, pp. 1--1, 2024.

\bibitem{manzoni2024integratedcommunicationimagingdesign}
\BIBentryALTinterwordspacing
M.~Manzoni, F.~Linsalata, M.~Magarini, and S.~Tebaldini, ``Integrated communication and imaging: Design, analysis, and performances of cosmic waveforms,'' 2024. [Online]. Available: \url{https://arxiv.org/abs/2405.19481}
\BIBentrySTDinterwordspacing

\bibitem{9468353}
D.~Solomitckii, M.~Heino, S.~Buddappagari, M.~A. Hein, and M.~Valkama, ``Radar scheme with raised reflector for nlos vehicle detection,'' \emph{IEEE Transactions on Intelligent Transportation Systems}, vol.~23, no.~7, pp. 9037--9045, 2022.

\bibitem{9547412}
S.~Wei, J.~Wei, X.~Liu, M.~Wang, S.~Liu, F.~Fan, X.~Zhang, J.~Shi, and G.~Cui, ``Nonline-of-sight 3-d imaging using millimeter-wave radar,'' \emph{IEEE Transactions on Geoscience and Remote Sensing}, vol.~60, pp. 1--18, 2022.

\bibitem{doi:10.1126/science.1210713}
\BIBentryALTinterwordspacing
N.~Yu, P.~Genevet, M.~A. Kats, F.~Aieta, J.-P. Tetienne, F.~Capasso, and Z.~Gaburro, ``Light propagation with phase discontinuities: Generalized laws of reflection and refraction,'' \emph{Science}, vol. 334, no. 6054, pp. 333--337, 2011. [Online]. Available: \url{https://www.science.org/doi/abs/10.1126/science.1210713}
\BIBentrySTDinterwordspacing

\bibitem{7109827}
G.~Oliveri, D.~H. Werner, and A.~Massa, ``Reconfigurable electromagnetics through metamaterials—a review,'' \emph{Proceedings of the IEEE}, vol. 103, no.~7, pp. 1034--1056, 2015.

\bibitem{9775078}
H.~Zhang, B.~Di, K.~Bian, Z.~Han, H.~V. Poor, and L.~Song, ``Toward ubiquitous sensing and localization with reconfigurable intelligent surfaces,'' \emph{Proceedings of the IEEE}, vol. 110, no.~9, pp. 1401--1422, 2022.

\bibitem{tagliaferri2023reconfigurable}
D.~Tagliaferri, M.~Mizmizi, G.~Oliveri, U.~Spagnolini, and A.~Massa, ``Reconfigurable and static em skins on vehicles for localization,'' \emph{IEEE Transactions on Wireless Communications}, pp. 1--1, 2024.

\bibitem{Buzzi_RISforradar_journal}
S.~Buzzi, E.~Grossi, M.~Lops, and L.~Venturino, ``{Foundations of MIMO Radar Detection Aided by Reconfigurable Intelligent Surfaces},'' \emph{IEEE Transactions on Signal Processing}, vol.~70, pp. 1749--1763, 2022.

\bibitem{9511765}
A.~Aubry, A.~De~Maio, and M.~Rosamilia, ``{RIS-Aided Radar Sensing in N-LOS Environment},'' in \emph{2021 IEEE 8th International Workshop on Metrology for AeroSpace (MetroAeroSpace)}, 2021, pp. 277--282.

\bibitem{9838638}
Z.~Wang, Z.~Liu, Y.~Shen, A.~Conti, and M.~Z. Win, ``{Source Localization with Intelligent Surfaces},'' in \emph{ICC 2022 - IEEE International Conference on Communications}, 2022, pp. 895--900.

\bibitem{9709801}
O.~Rinchi, A.~Elzanaty, and M.-S. Alouini, ``{Compressive Near-Field Localization for Multipath RIS-Aided Environments},'' \emph{IEEE Communications Letters}, vol.~26, no.~6, pp. 1268--1272, 2022.

\bibitem{9299878}
Y.~Tao and Z.~Zhang, ``Distributed computational imaging with reconfigurable intelligent surface,'' in \emph{2020 International Conference on Wireless Communications and Signal Processing (WCSP)}, 2020, pp. 448--454.

\bibitem{torcolacci2023holographic}
G.~Torcolacci, A.~Guerra, H.~Zhang, F.~Guidi, Q.~Yang, Y.~C. Eldar, and D.~Dardari, ``{Holographic Imaging with XL-MIMO and RIS: Illumination and Reflection Design},'' \emph{IEEE Journal of Selected Topics in Signal Processing}, pp. 1--16, 2024.

\bibitem{Alkhateeb23_imaging_comm}
H.~Luo and A.~Alkhateeb, ``Integrated imaging and communication with reconfigurable intelligent surfaces,'' in \emph{2023 57th Asilomar Conference on Signals, Systems, and Computers}, 2023, pp. 151--156.

\bibitem{bellini2024sensingnlosstroboscopicapproach}
\BIBentryALTinterwordspacing
D.~T. Bellini, D.~Tagliaferri, M.~Mizmizi, S.~Tebaldini, and U.~Spagnolini, ``{Sensing in NLOS: a Stroboscopic Approach},'' 2024. [Online]. Available: \url{https://arxiv.org/abs/2408.09883}
\BIBentrySTDinterwordspacing

\bibitem{10541333}
M.~Cui and L.~Dai, ``Near-field wideband beamforming for extremely large antenna arrays,'' \emph{IEEE Transactions on Wireless Communications}, pp. 1--1, 2024.

\bibitem{di2020smart}
M.~Di~Renzo, A.~Zappone, M.~Debbah, M.-S. Alouini, C.~Yuen, J.~De~Rosny, and S.~Tretyakov, ``Smart radio environments empowered by reconfigurable intelligent surfaces: How it works, state of research, and the road ahead,'' \emph{IEEE Journal on Selected Areas in Communications}, vol.~38, no.~11, pp. 2450--2525, 2020.

\end{thebibliography}

\end{document}